%% file: discussion.tex
\documentclass[a4paper]{article}

\usepackage{useful}
\usepackage[backend=bibtex,citestyle=authoryear-comp,%
  bibstyle=authoryear-comp,uniquename=false,useprefix=true,natbib=true,eprint=true]{biblatex}
\usepackage{graphicx}
\usepackage{epstopdf}

\usepackage{float}
\floatstyle{ruled}
\newfloat{code}{thp}{lop}
\floatname{code}{Code}

\title{Discussion of ``Geodesic Monte Carlo on Embedded Manifolds''}

\date{}
\begin{document}
\maketitle

\input{diaconis.tex}

\input{dryden.tex}
\input{kent.tex}
\input{pereyra.tex}

\input{shahbaba.tex}
\input{simpson.tex}
\input{rejoinder.tex}

\printbibliography

\end{document}

%% file: diaconis.tex
\section*{Comment: Connections and Extensions}
\subsection*{Persi Diaconis, Christof Seiler and Susan
  Holmes\footnote{Statistics Department, Stanford University, CA 94305}}

\subsubsection*{Historical Context}
We welcome this paper of Byrne and Girolami [BG]; it breathes even more life
into the emerging area of hybrid Monte Carlo Markov chains by introducing
original tools for dealing with Monte Carlo simulations on constrained spaces
such as manifolds.  We begin our comment with a bit of history. Using
geodesics to sample from the uniform distribution on Stiefel manifold was
proposed by \citet{Asimov} in his work on the Grand Tour for exploratory data
analysis.  For data $x_1,x_2,\ldots,x_n$ in ${\mathbb R}^p$, it is natural to
inspect low dimensional projections $\gamma x_1, \gamma x_2,\ldots, \gamma
x_n$ for $\gamma :{\mathbb R}^p\longrightarrow {\mathbb R}^k $.  In the [BG]
paper the authors have a space of k-frames in ${\mathbb R}^{p}$, called
$V_{k,p}$. If one chooses $\gamma$ at random from this space, the views would
be too `disconnected' or `jerky' for human observers. A better tactic turned
out to be to choose a few $\gamma_i, 1\leq i\leq L$, at random and then moving
smoothly from $\gamma_i$ to $\gamma_j$ by available closed form
geodesics. While in a historical mode, we point to the little known papers of
\citet{McLachlan2003} and more recent papers by Betancourt on hybrid
Monte Carlo \citep{betancourt2013}.
\subsubsection*{Discrete Hamiltonian Dynamics}

The paper of [BG] uses Hamiltonian dynamics to move around on a manifold in an intelligent way to get proposals for the Metropolis algorithm. There are also many problems where samples are needed for constrained discrete spaces. These
include sampling contingency tables with given row and column sums as in
\citet{DiaconisSturmfels}. We recently encountered the following problem in
a quantum physics context \citep{ChatterjeeDiaconis}. Consider boxes labeled ($1,2,3,\ldots)$. Drop $N$ balls into these boxes according to Bose-Einstein
allocation resulting in $N_i$ balls in box labeled $i$. Interest is on samples conditional on $\sum_{i} N_{i} i^{2} = E$. This is a discrete
version of the author's sampling from simplices and spheres. We do not currently have discrete versions of Hamiltonian dynamics apart from numerical schemes (leapfrog) that are used to solve the resulting differential equations as proposed by \citet{neal2011}. In contrast, [BG] compute the dynamics by splitting up the Hamiltonian into two analytically solvable parts.
We wonder whether the author's can suggest adaptions of their ideas to the discrete framework.

\subsubsection*{Non-Smooth Manifolds}
[BG] start with the Hausdorff measure from \textit{geometric measure theory} \citep{federer1969,morgan2009,diaconis2012} as a general way to define surface areas for non-smooth manifolds in arbitrary dimensions. We wonder if this is a bit misleading, since all subsequent developments and examples in the paper focus on homogeneous smooth manifolds. 

One example for which the methodology presented runs into difficulties is the barbell \citep{Grayson1989}, parametrized as:
\begin{equation*}
B\begin{pmatrix}x \\ \theta \end{pmatrix} = \begin{pmatrix} x \\ f(x) \cos(\theta) \\ f(x) \sin(\theta) \end{pmatrix}, \hspace{0.2cm} 0 \le \theta < 2\pi,
\end{equation*}
with changing radius:
\begin{equation*}
f(x) = 
\begin{cases}
r \cosh\left(\frac{|x|-l}{r}\right) & \text{if } |x| > l \\
r & \text{otherwise.}
\end{cases}
\end{equation*}
The difficulties arise at the corner of the transition from the bar to the bell section in the first coordinate of $B$ at position $|x| = l$. The derivative at these point is not defined.
In contrast, the geometric measure theory approach handles such difficulties by realizing that sets of area 0 do not influence the integral over a manifold. The intuition is that the line dividing the bar and the bell is a line which is negligible for computing two-dimensional integrals. 
Following this approach as described in \citet{diaconis2012}, we sample $x$ from the unnormalized surface measure:

\begin{equation*}
\sqrt{\operatorname{det} \left[ \operatorname{\bold{D}}B\begin{pmatrix}x \\ \theta \end{pmatrix} \right]^{T} \left[ \operatorname{\bold{D}}B\begin{pmatrix}x \\ \theta \end{pmatrix} \right] } = 
\begin{cases}
r \cosh^{2}\left(\frac{|x|-l}{r}\right) & \text{if } |x| > l \\
r & \text{otherwise.}
\end{cases}
\end{equation*}
The R code snippet (Code \ref{Code:RejectionSampling}) generates samples using rejection sampling for parameter $x$.
\begin{code}[b]
\begin{verbatim}
n = 5e3; r = 1; l = 2; L = 4
xprop = runif(n, min = -L, max = L)
eta = runif(n, min = 0, max = (r * cosh((abs(L) - l)/r)^2))
x = c()
for (i in 1:length(xprop)) {
    if (abs(xprop[i]) > l) {
        if (eta[i] < (r * cosh((abs(xprop[i]) - l)/r)^2)) {
            x = c(x, xprop[i])
        }
    } else {
        if (eta[i] < r) {
            x = c(x, xprop[i])  }}}
\end{verbatim}
\caption{Rejection sampling yielding $x$.}
\label{Code:RejectionSampling}
\end{code}

\begin{figure}[t]
\begin{center}
\includegraphics[width=1.0\textwidth]{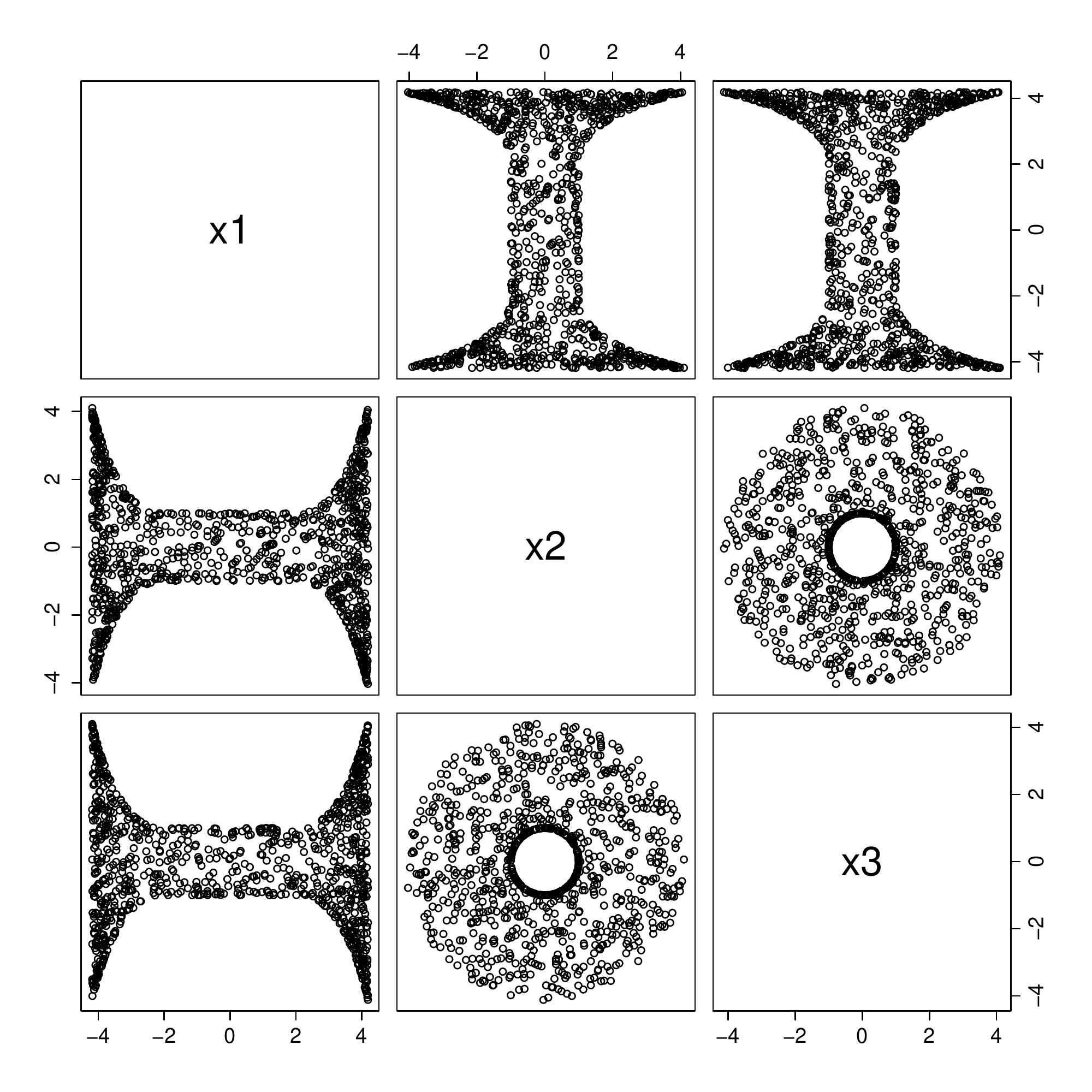}
\caption{The barbell is an example of a non-smooth manifold.}
\label{fig:BarbellUniformSufraceArea}
\end{center}
\end{figure}

From these samples, and $\theta$ drawn uniformly between $0$ and $2\pi$, we can plot the barbell with points uniformly distributed with respect to its surface area (Figure \ref{fig:BarbellUniformSufraceArea}). If we sampled points uniformly from the parameters $x$, we would obtain higher point density on the bar than on the bell section due to higher curvatures.

We are curious to know why [BG] decided to include the geometric measure theory part in the introduction and not to just simply focused on Riemannian manifolds and the Riemmanian volume form.

\subsubsection*{Consistency of Bayes Estimates on Manifolds.} 

Two different philosophical view points are crucial to study the consistency of Bayes estimates, namely ``classical'' and ``subjectivistic''.
The classical view point studies the consistency of Bayes estimates assuming the existence of a fixed underlying parameter. In this context, we consider the posterior Bayes estimate to be consistent w.r.t. a prior if it converges to the underlying parameter as the number of imaginary observations tends to infinity.

On the other hand, the subjectivistic view point is nihilistic of a fixed underlying parameter. In this context, we rather evaluate if two different priors created by two different imaginary statisticians converge to the same posterior estimate as the number of imaginary observations tends to infinity. 
We can analyze the derivative of the map that sends the prior to the posterior measure. This helps to evaluate how the posterior reacts to small changes in the prior. In this fashion, we can study an infinite amount of imaginary statisticians and how their beliefs affect the outcome of Bayesian analysis. 

We introduced these concept in \citet{DiaconisFreedman1986} for Euclidean spaces, and we are interested in how these results translate to the case of smooth and non-smooth manifolds. Some initial work towards addressing these questions can be found in \citet{Bhattacharya2012}.

\subsubsection*{Manifold and Metric Learning}
In the absence of a manifold parametrization we might want to estimate it from
data.  Recent advances by \citet{PerraulJoncas2013} on unifying manifold
learning methods into a consistent framework by learning the Riemannian metric
in addition to the manifold and its embedding are promising but build upon the
assumption of uniform sampling density on the manifold
\citep{VonLuxburg2008,Belkin2007}.  But what if the sampling of the data is not
related to the geometry of the manifold? In this case, we want to find the
manifold that is consistent for a family of distributions for a given set of
data points.  From a Bayesian perspective, we could study non-uniform density
distributions on the manifold through the derivative of the map from prior to
posterior measure analog to consistency evaluations of Bayes estimates.

\subsubsection*{Applications in Computational Anatomy}
Among many potential fields of application, we would like to highlight \textit{computational anatomy} \citep{Miller2004,Younes2010,Marsland2012}. 
The main goal of computational anatomy is to compare shapes of organs (e.g. brain, heart and spine) observed from computed tomography (CT) and magnetic resonance imaging (MRI). Statistical analysis of shape differences can be useful to understand disease related  changes of anatomical structures. The key idea is to estimate transformations between a template and patient anatomies. These transformation encode the structural differences in a population of patients. There is a wide range of groups of transformations that have been studied, ranging from rigid rotations to infinite dimensional groups of diffeomorphisms. What elements across groups have in common is that they do not live in Euclidean space but on more general manifolds. Currently, most transformation estimators are based on optimization of a cost function. In the future, we envision Bayesian approaches along the line of \cite{SeilerGSI2013} with the help of methodologies proposed in this paper.

\subsubsection*{Future Directions} 
The paper suggests new research questions: how long should the new algorithms be run to ensure that the resulting distributions are usefully close to their stationary distribution? We haven't seen any careful analysis of Hybrid Monte Carlo in continuous
problems (we mean quantitative, non-asymptotic bounds as in \citet{HorbertJones}).
A first effort was made in a toy problem in \citet{DiaconisHolmesNeal}.

The authors work with `nice manifolds', often manifolds are only given implicitly,
with local coordinate patches. Our work \citep{diaconis2012} did not deal with this problem, we would love to have help from the authors to make progress in these types of applications.


%% file: dryden.tex
\section*{Comment}
\subsection*{Ian L.~Dryden\footnote{School of Mathematical Sciences, University of Nottingham}}

The authors have introduced an interesting and mathematically intricate method for 
Markov chain Monte Carlo simulation on an embedded manifold. The geodesic Monte Carlo (MC) method provides large 
proposals as part of the scheme, which are devised by careful study of the 
Riemannian geometry of the space and the geodesics in 
particular. The aim of the resulting algorithm is to produce a chain with low 
autocorrelation and high acceptance probabilities. As displayed by the authors, 
the method is well geared up for simulating from unimodal distributions on a manifold 
via the gradient of the log-density and the geodesic flow. They also demonstrate its effective 
use in multimodal scenarios via parallel tempering. Given that there are always many choices of embedding, 
should one choose as low dimensional embedding as possible? 

There are various levels of approximation in the algorithm and so it is worth exploring in
any specific application if simpler algorithms can end up providing more efficient or more
accurate simulations. Consider the Fisher-Bingham example, and recall that the 
Fisher-Bingham $(c,A)$ distribution can be defined as 
$$\{ X | \|X\|=1 \} \; \; \; {\rm where} \; \; X \sim N_p( \mu , \Sigma ) ,  $$
with $\mu = -\frac{1}{2}(A+aI_p)^{-1}c, \Sigma = -\frac{1}{2}(A + aI_p)^{-1}$, 
$a$ is chosen such that $(A+aI_p)$ is negative definite \citep[see][p.175]{mardia2000} and $I_p$ is the $p \times p$ identity matrix. 
Since the Fisher-Bingham density is unchanged by adding $aI_p$ to $A$, we can, for example, 
choose $a$ such that ${\rm trace}(\Sigma) = 1$. 
The integrating constant of the Fisher-Bingham 
can be expressed in terms of the density of a linear combination of noncentral
$\chi^2_1$ random variables \citep{Kumewood05}, which can be evaluated using a saddlepoint approximation.
Hence simulation via rejection methods is feasible. 

\begin{figure}[htbp]
\begin{center}
\includegraphics[width=8cm,trim=100 100 100 50]{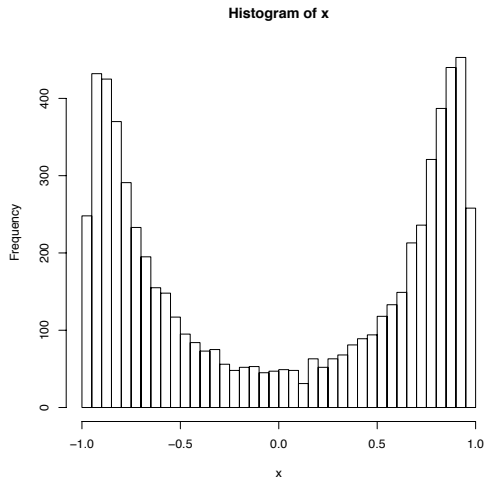}
\end{center}
\caption{Simulated values of $x_5$ for the Fisher-Bingham example with $c=0$. There are 6588 
simulated values from 2 billion proposals.}
\label{FIG1}
\end{figure}

An even simpler approach when $c$ is small could be to simulate from 
$Y \sim N_p( \mu , \Sigma )$, and then keep only the observations that fall within 
$| \|Y \|-1 | < \nu$, for small $\nu > 0$. This naive conditioning method 
might appear rather inefficient, but the accepted
observations are independent draws. Note that if the dimension 
$p$ is large and $X$ Bingham distributed with ${\rm trace}(\Sigma) = 1, {\rm trace}(\Sigma^2) \approx 1$, 
$c=0$ then from \citet{Dryden05} we have the approximation 
$X \approx N_p( 0 , \Sigma)$. Hence, even for large $p$ this can still be a practical 
method for certain $\Sigma$. 
In Figure \ref{FIG1} we show the results of this algorithm in the example from 
Section 5.1 of the paper, with $c=0$ and with 2 billion proposals and $\nu = 2 \times 10^{-6}$. Here 
$a=-23.06176$ 
and the acceptance rate is $0.00033\%$.    

There is always a trade-off with any simulation method, and one needs to compromise 
between the level of approximation (through $\nu$ here), the efficiency in run time, the independence 
of observations and the amount of coding involved in the implementation. 
For this Bingham example the naive conditional method 
seems reasonable here, giving independent, near exact realisations and very minimal effort in coding. 
However, the beauty of the geodesic MC method of the paper is that the algorithm is quite general, and so 
can be tried out in a range of scenarios where there may be no reasonable alternative.


%% file: kent.tex
\section*{Comment}
\subsection*{John T. Kent\footnote{Department of Statistics, University of Leeds, Leeds LS2 9JT, UK}}

Statistical distributions on manifolds have become an increasingly
important component of geometrically-motivated high-dimensional
sophisticated statistical models in recent years. For example,
\citet{green2006} used the matrix Fisher distribution for random
$3 \times 3$ rotation matrices as part of a high-dimensional Bayesian
model to align two unlabelled configurations of points in
$\mathbb{R}^3$, with an application to a problem of protein alignment
in bioinformatics.  MCMC simulations often form the standard
methodology for fitting such high-dimensional models.  Hence there is
a growing interest in developing efficient and general methods for
simulating distributions on manifolds in their own right.  The paper
makes a very valuable contribution in this area.

However, although MCMC is a very general and very powerful
methodology, it is inherently potentially slow and cumbersome to use
in practice, due to the formal need to run a Markov chain to
convergence.  Hence when quicker alternatives (such as acceptance
rejection algorithms) are available, it is important to be aware of
them.

Recent developments in acceptance rejection algorithms on spheres and
related manifolds have greatly increased the scope of acceptance
rejection methods for distributions such as Fisher, Bingham and
Fisher-Bingham.  The underlying idea is to use the angular central
Gaussian distribution (which is easy to simulate) as an envelope for a
Bingham distribution.  In turn the Bingham distribution can be used as
an envelope for the Fisher and Fisher-Bingham distributions.  The
basic idea works in all dimensions.  Further the efficiencies can
often be guaranteed to be very reasonable.

As an elegant application of this general methodology, consider the
matrix Fisher distribution on SO(3), the special orthogonal group of
$3 \times 3$ rotation matrices. This distribution is often used to
model unimodal behavior about a preferred rotation matrix.  There is
an elegant mathematical identity between $SO(3)$ and $S_3$, the unit
sphere in 4 dimensions, and it also follows that the matrix Fisher
distribution on $SO(3)$ can be identified with the Bingham
distribution on $S_3$.  Hence the new method for the Bingham
distribution can be used directly for the matrix Fisher in this
setting.  It can be shown that the efficiency of this new acceptance
rejection simulation method is very respectable; it is bounded below
by 45\% for all values of the parameters.  More details can be found
in \citet{kent2013}.

It must be conceded that this new acceptance rejection methodology is
not a panacea. In particular, for product manifolds there is often
currently no alternative to MCMC.  But for the simpler cases, the
acceptance rejection methods can be very effective.


%% file: pereyra.tex
\section*{Comment}
\subsection*{Marcelo Pereyra\footnote{Department of Mathematics, University of Bristol}}

I congratulate the authors for an interesting paper and an important methodological contribution to the problem of sampling from probability distributions on manifolds.  As an image processing researcher I shall restrict my comments to the potential of the proposed methodology for statistical signal and image processing. There are numerous new and exciting signal and image processing applications that require performing statistical inference on parameter spaces constrained to submanifolds of $\mathbb{R}^n$ and for which the proposed HMC algorithm is potentially interesting. For example, there are many \emph{unmixing} or \emph{source separation} problems that require estimating parameters that, because of physical considerations, are subject to positivity and sum-to-one constraints (i.e. constrained to a simplex) \citep{Golbabaee_2012a}. For instance, the estimation of abundances (or proportions) of different materials and substances within the pixels of a satellite hyperspectral image \citep{Bioucas_IEEE_JSTARS_2012}. These images are increasingly used in environmental sciences to monitor the evolution of vegetation in rainforests and in agriculture to forecast crop yield. Similar spectral imaging technologies are now used in material science and chemical analysis \citep{Dobigeon_Ultramicroscopy_2012}. Moreover, another important example of signal processing on manifolds is dictionary learning for sparse signal representation and compressed sensing, which involves estimating a set of orthonormal vectors constrained to a Stiefel manifold \citep{Dobigeon_IEEE_Trans_SP_2010}. Similar models arise in compressed sensing of low-rank matrices, which find applications in sensor networks and sparse principal component analysis \citep{Golbabaee_2012b}. The methodology presented in this paper is potentially very interesting for these and many other modern applications. However, in order for the proposed HMC algorithm to be widely adopted in signal processing it is fundamental to introduce efficient adaptation mechanisms to tune the HMC parameters automatically. I wonder whether the authors have considered an adaptive version of their algorithm, possibly by using an approach similar to the one recently presented in \citet{wang2013} for other HMC algorithms. The publication of an open-source MATLAB toolbox would also contribute greatly to its dissemination in the statistical signal and image processing communities.

Modern signal processing and machine learning applications have motivated the development of powerful new methods to perform statistical inference on high-dimensional manifolds. Most effort has been devoted to the development of new optimization methods that give access to maximum a posteriori estimates \citep{Combettes2011,Alfonso2012}. Sampling methods in general and the proposed HMC algorithm in particular can allow performing a significantly richer Bayesian analysis (i.e. they allow approximating expectations such as posterior moments, posterior probabilities or quantiles, and Bayesian factors useful for hypothesis testing and model choice). Therefore the methodology presented in this paper has the potential to not only impact the specific applications mentioned above, but to sustain and promote the adoption of Bayesian methods in general in signal and image processing. 

Finally, it would be interesting to explore connections between the proposed HMC algorithm and state-of-the-art optimisation methods for parameters constrained to manifolds \citep{Combettes2011,Alfonso2012}. A first step in this direction could be the paper I authored \citep{Pereyra2013} which highlights the great potential for synergy between MCMC and modern convex optimisation.


%% file: shahbaba.tex
\section*{Comment}
\subsection*{Babak Shahbaba\footnote{Department of Statistics and Department
    of Computer Science, University of California, Irvine, USA.}, Shiwei
  Lan\footnote{Department of Statistics, University of California, Irvine,
    USA.} and Jeffrey Streets\footnote{Department of Mathematics, University of California, Irvine, USA.} }

We would like to start by congratulating Byrne and Girolami for writing such a thoughtful and extremely interesting paper. This is in fact a worthy addition to other high impact papers recently published by Professor Griolami's lab in this field. The common theme of these papers is to use geometrically motivated methods to improve efficiency of sampling algorithms. In their seminal paper, \citet{girolami2011} propose a novel HMC method, called Riemannian Manifold Hamiltonian Monte Carlo (RMHMC), that adapts to the local geometry of the parameter space. While this is a natural and beautiful idea, there are significant computational difficulties which arise in effectively implementing this algorithm.   In contrast, in this current contribution, Byrne and Girolami focus on special probability distributions which give rise to particularly nice Riemannian geometries. In particular, the examples under consideration described in section 4 allow for closed-form
solutions to the geodesic equation, which can be used to reduce computational cost of geometrically motivated Monte Carlo methods. 

While the proposed splitting algorithm is quiet interesting, we initially doubted its impact since Riemannian metrics with closed-form geodesics are extremely rare. However, we are now convinced that this approach will likely see application beyond what is outlined herein. For example, we believe that this approach can be used to improve computational efficiency of sampling algorithms when the parameter space is constrained. The standard HMC algorithm needs to evaluate each proposal to ensure it is within the boundaries imposed by the constraints. Alternatively, as discussed by \citet{neal2011}, one could modify standard HMC so the sampler bounces back after hitting the boundaries. In Appendix A, Byrne and Girolami discuss this approach for geodesic updates on the simplex. 

\begin{figure}
\begin{center}
\centerline{\includegraphics[width=4in, height=1.9in]{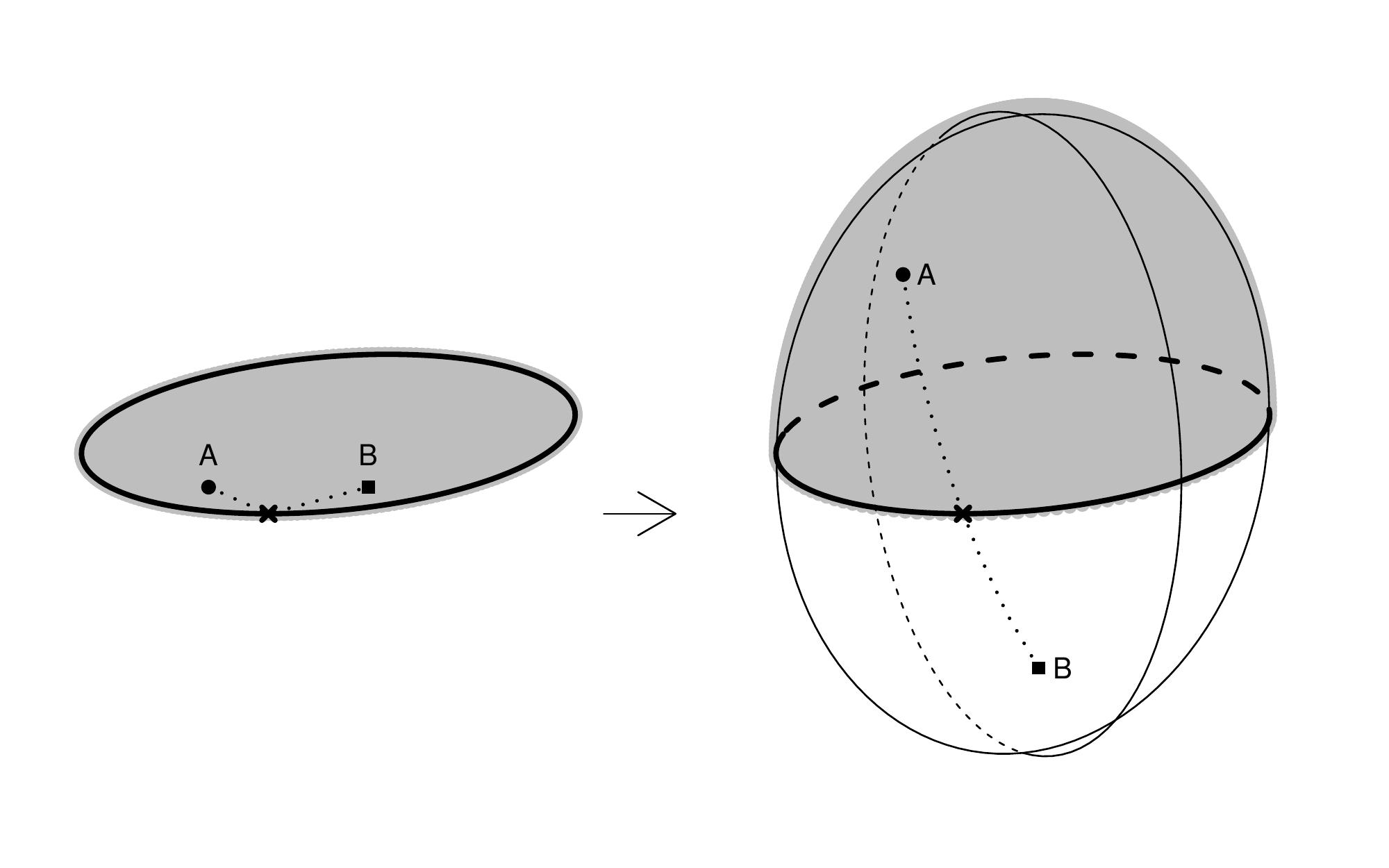}}
\caption{Transforming unit ball ${\bf B}_0^D(1)$ to sphere ${\bf S}^D$.}
\vspace{-15pt}
\label{fig:B2S}
\end{center}
\end{figure}

In many cases, a constrained parameter space can be bijectively mapped to a unit ball, ${\bf B}_0^D(1):=\{\theta\in\mathbb R^D: \Vert \theta\Vert_2 =\sqrt{\sum_{i=1}^D \theta_i^2}\leq 1\}$. Augmenting the parameter space with an extra auxiliary variable
$\theta_{D+1} = \sqrt{1-\Vert \theta\Vert_2^2}$, we could form an extended parameter space, $\tilde \theta = (\theta, \theta_{D+1})$ so that the domain of the target distribution 
changes from unit ball ${\bf B}_0^D(1)$ to \emph{$D$-Sphere} ${\bf S}^D =\{\tilde \theta\in
\mathbb R^{D+1}: \Vert \tilde\theta\Vert_2=1\}$, 
\begin{equation}\label{b2s}
T_{{\bf B}\to {\bf S}}: {\bf B}_0^D(1)\longrightarrow {\bf S}^D, \quad \theta \mapsto \tilde\theta = (\theta, \pm\sqrt{1-\Vert \theta\Vert_2^2})
\end{equation}
Sampling from the distribution of $\tilde\theta$ on ${\bf S}^D$ can be done efficiently using the Geodesic Monte Carlo approach, which allows the sampler to move freely on ${\bf S}^D$, while its projection onto the original space always remains within the boundary. This way, passing across the equator from one hemisphere to the other will be equivalent to reflecting off the boundaries as shown in Figure \ref{fig:B2S}. 

Our last comment is related to the embedding procedure discussed in Section 3.2. We wonder if such embedding and the resulting extra step for projection could be avoided by writing the dynamics in terms of $(q, v)$ in the first place and splitting it as follows: 
\begin{eqnarray}
\left\{\begin{array}{lcl}
\dot q & = & 0\\
\dot v & = &  G^{-1} \nabla \log \pi_{\mathcal H}(q)
\end{array}\right.\label{LD:U}
\qquad
\left\{\begin{array}{lcl}
\dot q & = & v\\
\dot v & = & -v^T \Gamma v
\end{array}\right.\label{LD:K}
\end{eqnarray}
where $\Gamma$ is the Christoffel symbol of second kind. The second dynamics in \eqref{LD:K} is regarded as the general geodesic equation:
\begin{equation}
\ddot q + \dot q^T \Gamma \dot q = 0
\end{equation}
The first dynamics in \eqref{LD:U} is solved in terms of $(q,v)$ in a more natural way:
\begin{equation}
q(t) = q(0) \quad\textrm{and}\quad v(t) = v(0) + tG(q)^{-1}\nabla_{q} \log \pi_{\mathcal H}(q) \big|_{q=q(0)}
\end{equation}
This way, we avoid the additional projection step and have $v(t)\in T_{q(t)}\mathcal M$ as long as $v(0)\in T_{q(0)}\mathcal M$. This also serves to isolate what seems to be the key point in this work, which is not that the dynamics are taking place on an embedded manifold, but that they are taking place on a manifold \emph{whose geodesics are known explicitly}. With this viewpoint the applicability of the ideas of this paper should be further expanded.


%% file: simpson.tex
\section*{Comment}
\subsection*{Daniel Simpson\footnote{Department of Mathematical Sciences,
    Norwegian University of Science and Technology, N-7491 Trondheim, Norway. Email: \texttt{Daniel.Simpson@math.ntnu.no}}}

The basic idea of simulation-based inference is that we can approximately
calculate anything we like about a probability distribution if we can draw
independent samples from it.  This means that we can use sampling to explore
the posterior distribution and it turns out that the quantities we compute
will usually have an error of $\mathcal{O}(N^{-1/2})$ if they are calculated
from $N$ samples. Unfortunately, in almost any realistic situation, we cannot
directly simulate from the posterior, however the remarkable (and their
ubiquity really shouldn't detract from just how remarkable MCMC methods are)
Markov Chain Monte Carlo idea says that it's enough to take a chain of
dependent simulations that are heading towards the posterior distribution and
use these simulations to calculate any quantities of interest. The variance in
the estimators still decay like $\mathcal{O}(N^{-1/2})$ and they pretty much
always work eventually.  (There is, of course, an entire world of details
being suppressed within the world `eventually'.)

The problem with vanilla (Metropolis Hastings) MCMC methods is that they are
slow.  It's fairly easy to see why this is true: whereas perfect Monte Carlo
methods `know' enough about the posterior to produce perfect samples,
Metropolis Hastings algorithms only require the ability to calculate ratios of
the posterior density.  For simple models, this may not be a problem, but as
the posterior distribution becomes more complicated, it's fairly
straightforward to imagine the the efficiency of schemes based on simple
proposals will plummet.  Byrne and Girolami consider the even more complicated
situation where the natural parameters of the model have a non-linear
structure. These type of models arise frequently in ecology.  A simple example
occurs when modelling community structure in ecology, in which case the
association between the occurrence of different species is modelled as a
symmetric positive definite matrix \citep{ovaskainen2011making}.  A more
complicated example occurs in paleoclimate reconstruction, where one is often
required to model `compositional data', that is proportions (rather than
counts) of different types of pollen in a sample \citep{salter2006modelling}.
A simple model for proportions is the Dirichlet distribution, however this is
frequently unsuitable due to real compositional data having a large number of
zero proportions.  More complicated distributions for proportions can be
written as distributions on a simplexes, which are considered by
Byrne and Girolami.  In these situations, it is often not even obvious how to
construct {bad} proposals, let alone efficient ones!

Typically, however, we know a lot about the model that we are trying to infer.
In this case, it makes sense to include all of the information that we have in
order to make the MCMC algorithm explore the posterior in a more efficient
manner.  In particular, people working within well understood statistical
frameworks, such as modelling with latent Gaussian models, have been able to
use analytical results to design MCMC schemes \citep{art192,art412} or other
approximate inference methods \citep{art451}.  For more general statistical
models, \citet{girolami2011} constructed a general framework for
constructing efficient MCMC schemes based on the classical links between
statistical modelling and differential geometry.

The innovation of \citet{girolami2011} is to provide an essentially
automatic way to improve MCMC performance by using standard concepts from
statistical asymptotics. The idea is that, even if we don't know everything we
would like to know about the posterior distribution, we can approximate what
it's like ``on average''.  Specifically, this means that we can, for each
point in the parameter space, find a Gaussian distribution that locally looks
like an average posterior (where the average is taken over the data).  We can
then construct a proposal distribution based on this approximation and it is
reasonable to expect it to perform better than a naive choice.
\citeauthor{girolami2011} proposed two basic types of algorithm: The
first was a version of Metropolis-adjusted Langevin algorithm (MALA) uses this
approximation to propose a new value that's nearby the current point, while
the second algorithm is version of Hamiltonian Monte Carlo (HMC) chains
together a number of these local approximations to try to make a proposal in a
distant part of the parameter space.  As such, one expects HMC to be more
statistically efficient (that is, the samples are less dependent), while the
MALA proposals are more computationally efficient (that is, they take less
time to compute).

The method described by \citet{girolami2011} is more general than the one described above.  Their framework, which is described in the language of differential geometry, allows for almost any type of local second-order structure.  For common problems, where the parameter space is $\mathbb{R}^d$, the only requirement is that each point in the parameter space is associated in a smooth way with a symmetric positive definite matrix. In this case, it makes sense for these matrices to be built from local approximations to the posterior distribution and the whole scheme can be easily described without ever appealing to the slightly intimidating notion of a manifold.  

The case considered by Byrne and Girolami is different.  Here the parameter space isn't flat and the notion of a manifold becomes essential to defining good inference schemes.  The methods considered by Byrne and Girolami are different from the geometrically simpler models considered by \citet{girolami2011}.  Rather than introducing a geometric structure in order to better explore a distribution on $\mathbb{R}^n$, Byrne and Girolami use the \emph{natural} geometry of the parameter space to construct a proposal.  It is unsurprising that this strategy results in efficient MCMC schemes: it is almost universally true that numerical methods that are consistent with the underlying structure of the problem are more efficient than those that aren't!  

That is not to say that the extra efficiency from using the problems natural manifold structure comes for free.  Hamiltonian Monte Carlo methods are based on the approximate integration of Hamilton's equations, which are symplectic ordinary differential equations in position and momentum space.  Integrating symplectic ODEs is an active field of research and actually implementing these integrators can be quite challenging.  In particular, the HMC method proposed by \citet{girolami2011} requires, at each step, the solution of a non-linear system of equations, which can cause the manifold HMC proposal to catastrophically fail if it is programmed incorrectly.  Fortunately, Byrne and Girolami show that when the parameter space is an embedded manifold, it is possible to use a much simpler integrator. In order for their splitting technique to be applicable, it is necessary to have an explicit expression for geodesic flow on the parameter manifold and, in the cases considered in the paper, this exists. Given an explicit form of the geodesic, one only has two choices left: the step size $\epsilon$ and the number of steps $N$ in each proposal.  The performance of HMC methods are known to be very sensitive to these parameters, however recent advances in (non-manifold) HMC suggests that it is possible to adaptively select these in an  efficient manner \citep{hoffman2013}.

As Byrne and Girolami have focused on building HMC methods on embedded manifolds, it is instructive to examine the barriers to similarly generalising the manifold MALA schemes.  Recall that MALA-type methods on $\mathbb{R}^n$ are biased random walks that propose a new value  $\theta^*$ by as $$\theta^* - \theta^{(k)} \sim N(\mu( \theta^{(k)}), H( \theta^{(k)})^{-1}),$$ where the specific forms of $\mu(\cdot)$ and $H(\cdot)$ are irrelevant to this discussion.  The problem with generalising this type of proposal to a manifold is obvious: the subtraction operation does not make sense.  One way around this problem is to take a lesson from the optimisation literature and note that we can make sense of this proposal using tangent spaces and exponential mappings (or, more generally, retractions)\citep{absil2009optimization}.  In this case, we propose $$
\theta^* = R_{\theta^{(k)}}(p^{(k)}),
$$ where $R_{\theta^{(k)}}(\cdot) : T_{\theta^{(k)}}\mathcal{M} \rightarrow \mathcal{M}$ is a retraction map and $p^{(k)} \sim N(\mu( \theta^{(k)}), H( \theta^{(k)})^{-1})$ is a random vector in the tangent space $T_{\theta^{(k)}}\mathcal{M}$ \citep{absil2009optimization}.  The problem with this proposal mechanism is that it is not obvious how to compute the proposal density, which is required when computing the acceptance probability.   Hence, there is no clear way to design a MALA-type scheme that respects the non-linear structure of the parameter space.


%% file: rejoinder.tex
\section*{Rejoinder}
\subsection*{Simon Byrne and Mark Girolami\footnote{Department of Statistical
    Science, University College London}}

We would like to thank all respondents for their interesting comments, which
clearly identify exciting areas for further investigation.

Both Kent and Dryden highlight recent developments in rejection sampling
methods for obtaining independent samples form distributions on
manifolds. Such methods are obviously preferable when available, however as
mentioned in section 5.1, the danger being is that rejection-based techniques
can have exponentially low acceptance rates, particularly in
higher-dimensional problems. Indeed the impressive results of Kent, Ganeiber,
and Mardia in avoiding this problem by obtaining constant lower-bounds of the
acceptance rates highlights the importance of considering the underlying
geometry of the manifold.

Pereyra and Simpson point out the many links with optimisation: indeed
optimisation over manifolds has a rich history, and there is a wealth of
literature with many interesting algorithms. However, as Simpson points out,
many of these algorithms are based on projection operators, and thus we face
what could be described as the "Curse of Detailed Balance": the difficulty of
computing of the reverse proposal, which is required for the evaluation of the
acceptance ratio to ensure we are targeting the correct invariant
density. Hamiltonian-based methods are able to exploit symplectic geometric
structure---namely reversibility and volume preservation---in a manner that
makes this almost trivial,

We are very excited to see that Shahbaba, Lan and Streets have had success
with this methods. We agree entirely with their point that it is the explicit
geodesics, and not the embedding, which makes this method successful: our
reason for using the embeddings is that in all cases we identified, the
embeddings proved convenient to work with. Our reason for using the projection
is that this is typically of lower computational cost than inversion of $G$.

As several commenters point out, despite its long history, remarkably little
is known about the theoretical properties of the HMC algorithm, especially
when compared to say Gibbs sampling and Metropolis--Hastings algorithms based
on random-walks and Langevin diffusions. In particular one open question is
the optimal tuning of the step-size and integration length
parameters. Unfortunately HMC is not readily amenable to the usual
probabilistic tools, such as links to diffusions, due to the precise property
that makes it so powerful: the ability to simulate long trajectories and make
distant proposals. This is an open question, attracting interest from numerous
researchers.

The paper by \citet{wang2013} propose an empirical Bayesian optimisation
approach, but this comes with significant overhead in obtaining sufficient
samples on which to base the objective function, and provides little insight
into theoretical behaviour. We think that future advances will perhaps require
a larger set of tools, such as exploiting the rich geometric structure and
elegant numerical properties of Hamiltonian
methods \citet[\eg][]{hairer2006}. This is already an area of active
research, for instance the recent work of \citet{beskos2013} utilises the
tools of backward error analysis to obtain an asymptotic-in-dimension bound of
the optimal acceptance rate. Other recent advances are the "no U-turn"
approach of \citet{hoffman2013}, which seeks to truncate the integration
path based on a geometric criterion, and the general-purpose SoftAbs metric
of \citet{betancourt2013} for RMHMC. Nevertheless, there are many
interesting open questions in this area, which we intend to pursue further.

As Pereyra points out, the development of software toolboxes will greatly
lower the barrier to implementation, enhancing the utility of these
methods. Indeed, the venerable BUGS software and its descendants have
revolutionised applied Bayesian statistics over the past twenty years. The
rapidly-developing STAN library \citep{stan}, aims to do the same using
HMC, incorporating tools such as automatic differentiation to simplify the interace,
and its impressive early results seem set to make it the heir-apparent to
BUGS. As we mention in the section on product manifolds, our methods dovetail
elegantly within a larger HMC scheme, and so would be a natural fit for such
software.

Although we derived Geodesic Monte Carlo in terms of smooth manifolds, it can
be easily extended to manifolds made of smooth patches, such as the barbell
example proposed by Diaconis, Seiler and Holmes, by appropriately modifying
the direction of the particle whenever it passes the boundary, in a similar
manner to the reflections used to constrain the particle to the simplex. Of
course this requires an explicit form of the geodesic of each patch, as well
as computing the point at which the particle crossed the boundary. A more
desirable approach would be to transform the space to a smooth manifold,
ideally preserving the topology, for instance the barbell could be transformed
into a cylinder.

One great challenge is extending HMC beyond Euclidean spaces. As mentioned by
Diaconis et. al., there is not an obvious analogue of HMC for discrete
spaces. In certain circumstances, it can be possible to augment the space with
additional continuous variables, which can allow the discrete variables to be
easily marginalised out, for example \citet{zhang2012} use a
Hubbard--Stratonovich transformation to apply HMC to the Ising model. Diaconis
et. al. also mention infinite-dimensional spaces such as diffeomorphism
groups: \citet{beskos2011} has demonstrated that HMC can be defined and
implemented for Hilbert spaces, and it would be exciting, both from a
theoretical and a numerical perspective, to extend it to yet more general
spaces.

These many open research questions will no doubt be developed in the coming
years, both theoretical analysis, methodological development and applications
to significant new and exciting areas.
